\documentclass[a4paper,11pt]{article}
\pdfoutput=1  % submitting via pdflatex (images in pdf/png/jpg)

\usepackage[T1]{fontenc}
\usepackage{amsmath}
\usepackage{amssymb}
\usepackage{graphicx}
\usepackage{multirow}
\usepackage{booktabs}
\usepackage{slashed}
\usepackage{gensymb}
\usepackage{jheppub}      
\usepackage{orcidlink}
\usepackage{float}  

\newcommand {\be}{\begin{equation}}
		\newcommand {\ee}{\end{equation}}
\newcommand {\ba}{\begin{eqnarray}}
		\newcommand {\ea}{\end{eqnarray}}

\def\e6{E(6)}
\def\10{SO(10)}
\def\21{SA(2) $\otimes$ U(1) }
\def\321{$\mathrm{SU(3) \otimes SU(2) \otimes U(1)}$ }

\def\422{SA(4) $\otimes$ SA(2) $\otimes$ SA(2)}

\def\roughly#1{\mathrel{\raise.3ex\hbox{$#1$\kern-.75em
			\lower1ex\hbox{$\sim$}}}} \def\lsim{\roughly<}
\def\gsim{\roughly>}

\def\lsim{\raise0.3ex\hbox{$\;<$\kern-0.75em\raise-1.1ex\hbox{$\sim\;$}}}
\def\gsim{\raise0.3ex\hbox{$\;>$\kern-0.75em\raise-1.1ex\hbox{$\sim\;$}}}

\def\roughly#1{\mathrel{\raise.3ex\hbox{$#1$\kern-.75em
			\lower1ex\hbox{$\sim$}}}} \def\lsim{\roughly<}
\def\gsim{\roughly>}

\def\lsim{\raise0.3ex\hbox{$\;<$\kern-0.75em\raise-1.1ex\hbox{$\sim\;$}}}
\def\gsim{\raise0.3ex\hbox{$\;>$\kern-0.75em\raise-1.1ex\hbox{$\sim\;$}}}

\title{\boldmath A $U(1)_{(B-L)_3}$ model for dark matter and \\
  the $B^+\to K^+\nu\bar{\nu}$ excess reported by Belle~II}

\author[a,b]{MohammadAli Abri}
\emailAdd{abri.mohammadali@ut.ac.ir}
\author[b]{Yasaman Farzan}
\emailAdd{yasaman@theory.ipm.ac.ir}
\affiliation[a]{Department of Physics, University of Tehran,\\
	P.O.Box 14395-547, Tehran, Iran}
\affiliation[b]{School of Physics, Institute for Research in Fundamental Sciences (IPM),\\
  P.O.Box 19395-5531, Tehran, Iran}

\abstract{We explore the $U(1)_{(B-L)_3}$ gauge symmetry under which only third generation fermions are charged. The cancellation of the $[U(1)_{(B-L)_3}]^3$ anomaly requires a chiral fermion, $\chi_L$, singlet under the gauge group but charged under $U(1)_{(B-L)_3}$. This chiral fermion can play the role of thermal dark matter with a relic abundance set via the freeze-out scenario through the annihilation to pairs of the third generation fermions.
In the quark mass basis, the new gauge boson will have off-diagonal couplings to the $b$ and $s$ quarks which can lead to a new contribution to $B^+\to K^+\nu_\tau\bar{\nu}_\tau$.
We discuss the relevant bounds from the LHC searches for a new gauge boson, dark matter searches, electroweak precision data, $\Upsilon$ decay, the CKM matrix and the $B_s^0-\bar{B}_s^0$ mixing. We entertain the possibility of explaining the $B^+\to K^+\nu\bar{\nu}$ excess recently reported by Belle~II within this model.}


\begin{document}
\maketitle
\flushbottom

\section{Introduction}
Although strong hints for the existence of dark matter (DM) in the universe exist, its nature is still unknown, opening up the window to build a plethora of models for the dark matter. A popular scheme for classifying the dark matter models is based on its portal to the Standard Model (SM) particles. The following three portals have received particular attention: (1)  Neutrino portal; (2) Higgs portal; (3) dark photon portal. Considering that the data  on the third generation of the quarks and leptons is limited, it is conceivable that the first hints of new physics may show up in the future precision measurements involving the third generation fermions. Bearing this in mind,  a model for dark matter is introduced in which the connection of DM to SM particles is through the third generation \cite{Hooper:2014fda,Blanco:2019hah,Kamada:2018kmi}. Taking the SM as a guiding star, this approach makes sense. In the SM, the third generation is also special as  the relatively large Yukawa couplings of the third generation break the global $U_L(3)
	\times U_R(3)$ flavour symmetry of  the SM gauge interactions to  $
	U_L(2)
	\times U_R(2)$.
%This scenario can lead to a new contribution to $B \to K^{(*)} \nu_\tau \bar{\nu}_\tau,  K^{(*)} \tau \bar{\tau}$.

In this paper, we gauge the  $B_3-L_\tau$ $U(1)$  symmetry under which  the flavour eigenstates $b$ and $t$ quarks have a charge of $1/3$ and the third generation leptons, $\tau$ and $\nu_\tau$, have a charge of $-1$. The rest of the SM fermions are neutral under this symmetry. We denote this symmetry by $U(1)_{(B-L)_3}$ and
show its  gauge boson with $V_\mu$.
We also  introduce a chiral fermion, $\chi_L$, singlet under the SM gauge group but charged under this new $U(1)$ which plays the role of the dark matter.

At one loop level, a kinetic mixing between $V_\mu$ and the hypercharge gauge boson appears, through which dark matter can scatter off nuclei. The direct dark matter searches then set a strong bound on the coupling, requiring that the dark matter to be of  Majorana nature which leads to the suppression of  the scattering cross section by the square of the dark matter velocity relative to the nucleus, $v_\chi^2$.  The annihilation of dark matter pair will  dominantly be $p$-wave, too, implying that the lower  bounds on the mass of thermal dark matter
 \cite{Fermi-LAT:2025gei} from the indirect searches for dark matter in the dwarf galaxies, the dark matter halo or the galaxy clusters can be relaxed; however, the dark matter annihilation in the DM spike around the central supermassive  black hole may  have observable effects \cite{Gondolo:1999ef,Balaji:2023hmy}. In the early universe, dark matter pairs can annihilate into pairs of third generation fermions through $s$-channel $V$ exchange, $\chi_L \bar{\chi}_L\to V^*\to t\bar{t},b\bar{b},\tau \bar{\tau},\nu_\tau\bar{\nu}_\tau$. Using micrOMEGAs7 \cite{Belanger:2026asz}, we determine the gauge coupling versus the gauge boson mass for which the observed relic DM abundance is obtained via the freeze-out scenario, taking into account the latest bounds from direct dark matter searches.

Since the $U(1)_{(B-L)_3}$ charges of the third generation quarks are different from those of the first and second generations, reproducing the mixings of the third generation quarks with the rest requires spontaneous symmetry breaking  through the vacuum expectation value (VEV) of a new Higgs boson charged under $U(1)_{(B-L)_3}$.
In the quark mass basis, there will be off-diagonal couplings of $V$ to the quarks which can give a contribution to the flavour changing neutral current (FCNC) processes such as $B^+\to K^+ \nu_\tau \bar{\nu}_\tau$. We shall discuss in detail whether this model can also explain the 2.7~$\sigma$ excess of
$Br(B^+\to K^+ \nu\bar{\nu})$ relative to the SM reported by the Belle~II collaboration \cite{Belle-II:2023esi}.

This paper is organised as follows. In sect.~\ref{model},  we review  the model and the bounds from  electroweak precision data, $\Upsilon$-decay, LHC search for extra heavy gauge  boson and the direct dark matter searches. In sect. \ref{sec:CKM}, we rewrite the gauge interaction in the quark mass basis, incorporating the information from the CKM mixing matrix and  discussing the bounds from the $B_s^0-\bar{B}_s^0$ and $B_d^0-\bar{B}_d^0$ mixings. In sect. \ref{SinglePHI},
we show how the CKM matrix elements can be reproduced in a minimal model with the VEV of  a heavy Higgs doublet charged under $U(1)_{(B-L)_3}$ with a relatively small VEV of $\sim 7$ GeV.
In sect. \ref{SymPHI}, we extend the model to include two such doublets, $\Phi$ and $\Phi'$, with a symmetry under $\Phi \leftrightarrow \Phi'$ and show that how within this model the stringent  bounds from $B_s^0-\bar{B}_s^0$ can be relaxed. In sect. \ref{sec:B+}, we discuss the contribution from our model to the  ${\rm Br}(B^+\to K^+ \nu \bar{\nu})$ excess reported by Belle~II. In sect. \ref{sec:DM}, we discuss the abundance of dark matter and the possibility of direct and indirect detection as well as the possibility of explaining the reported $B^+\to K^+ \nu \bar{\nu}$ excess. Our results are summarized in sect. \ref{sec:Con}.

%%%%%%%%%%%%%%%%%%%%%%%%%%%%%%%%%%%%%%%%%%%%%%%
\section{ The $U(1)_{(B-L)_3}$ model  for dark matter \label{model}}
%%%%%%%%%%%%%%%%%%%%%%%%%%%%%%%%%%%%%%%%%%%%%%%%%%%%%%
In this section, we will first review the Lagrangian and basics of the $U(1)_{(B-L)_3}$  model for dark matter.  We then review the bounds from the precision electroweak data, direct dark matter search experiments, $\Upsilon$ decay,  and  from collider searches for the gauge boson.  In sect.~\ref{sec:CKM}, we shall discuss the form of gauge interaction in the quark  mass basis as well as the relevant constraints from the CKM matrix. We will then derive   bounds on the off-diagonal couplings of $V$ to the quarks from the $B_s^0-\bar{B}^0_s
$ and $B_d^0-\bar{B}^0_d$ mixings.
In sect.~\ref{SinglePHI}, we introduce a minimal model with a single $\Phi$ which can reproduce the CKM matrix but is severely constrained by the $B_s^0-\bar{B}_s^0$ mixing. In sect.~\ref{SymPHI}, we introduce another variant of the model with $\Phi$ and $\Phi'$ symmetric under $\Phi \leftrightarrow \Phi'$ and show that in this  case, the bounds from  the $B_s^0-\bar{B}_s^0$ mixing can be relaxed, opening the room for explaining the $B^+\to K^+\nu \bar{\nu}$ excess.

Under  the  $U(1)_{(B-L)_3}$   symmetry,  while the third generation flavour  eigenstates, ({\it i.e.,} the $b$ and $t$ quarks) have a charge of $1/3$,  the  $U(1)_{(B-L)_3}$  charges of  $\tau$ and $\nu_\tau$ are equal to $-1$. The rest of the SM fermions are neutral under this symmetry.
We show the  gauge boson  of the  $U(1)_{(B-L)_3}$   symmetry with $V_\mu$.
Let us also  introduce a new chiral fermion, $\chi_L$,  singlet under the SM gauge group but charged under $U(1)_{(B-L)_3}$. In our model,  $\chi_L$ plays the role of the dark matter, stabilized by a $Z_2$ symmetry under which only $\chi_L$ is odd. The gauge interaction of the fermions can be written as follows:
\begin{equation} \label{gV/3}
	g_VV_\mu\left( \frac{\bar{t}\gamma^\mu t+\bar{b}\gamma^\mu b}{3}-\bar{\tau}\gamma^\mu\tau -\bar{\nu}_\tau\bar{\sigma}^\mu\nu_\tau + \bar{\chi}_L\bar{\sigma}^\mu \chi_L  \right),
\end{equation}
where $\bar{\sigma}^\mu=(1_{2\times 2},-\sigma^i)$ in which $1_{2\times 2}$ is the two by two identity matrix and $\sigma^i$ are the Pauli matrices.  The contribution from chiral $\chi_L$  helps to cancel
the $U(1)^3$ anomaly.
This new fermion should be much heavier than $O({\rm MeV})$ otherwise it will come to thermal equilibrium with $\nu_\tau$ for $T\sim$ MeV and will lead to $\Delta N_{eff}=1$ at the time of big bang nucleosynthesis which is ruled out. In particular, it cannot be identified with the light right-handed neutrino.
In the early universe, $\chi_L$ will be thermalised via its gauge interaction with the third generation fermions of the SM. As we shall discuss, the dark matter production mechanism can be the canonical freeze-out.

Like any new $U(1)$ gauge boson, $V_\mu$ can have a kinetic mixing with the hypercharge gauge boson, $B_\mu$, as
\begin{equation}
	-\frac{\epsilon}{2} (\partial_\mu V_\nu-\partial_\nu V_\mu)(\partial^\mu B^\nu-\partial^\nu B^\mu) \ .
\end{equation}
Since in this model the third generation is charged both under the
hypercharge $U(1)$ and the new $U(1)$, there will be a one loop contribution to the kinetic mixing  \cite{Kamada:2018kmi}:
\begin{equation}
	\epsilon=\frac{2 g_Y g_V}{9 \pi^2}\log \frac{\Lambda}{\mu}\simeq 0.02 \frac{g_V}{0.35}\log\left[\frac{\Lambda}{100~{\rm TeV}}\frac{100~{\rm GeV}}{\mu}\right]\ ,
\end{equation}
where $\Lambda$ and $\mu$ are respectively the cutoff of the integration of the loop and the energy scale of the experiment. There can be a tree level bare kinetic mixing which may partially cancel the contribution from the loop or add up to it. The total kinetic mixing (tree level plus the loop contribution) is constrained by the electroweak precision measurements. For $m_V\to m_Z$ the bounds can be very severe. However for $m_V>200$ GeV, the bound is about
0.1 \cite{Hook:2010tw,Ellis:2018xal} so the electroweak precision bound can be satisfied without a fine tuned cancellation between the loop level and tree level kinetic mixing.

%Moreover, since $\Phi$ is charged both under hypercharge and new gauge symmetry, its %VEV can induce a mass mixing between $V$ and the Z boson \cite{Ellis:2018xal}:
%$$\delta m^2=\frac{e g_V}{6 \sin \theta_W \cos\theta_W}\langle \Phi\rangle^2\sim 0.22 ~{\rm GeV}^2\frac{g_V}{0.35}\left(\frac{\langle \Phi\rangle}{4 ~{\rm GeV}}\right)^2 \ll  m_Z^2\sin \epsilon $$ so the effect of $\delta m^2$ on the oblique parameters will be much smaller than that of $\epsilon$. 

In our model in which the dark matter also couples to $V_\mu$, the kinetic mixing can lead to the scattering of dark matter off nuclei and therefore strong bounds on the kinetic mixing, $\epsilon$.  Dismissing a fine tuned cancellation between a tree level and loop level contributions to $\epsilon$, the upper bound on  $\epsilon$ can be translated into an upper bound on the order of magnitude of $g_V/m_V$.
As shown in Ref. \cite{Bell:2014tta}, if $\chi_L$ is of Dirac type, the bounds from the direct dark matter searches rule out values of $g_V$ and $m_V$ for which  the freeze-out scenario can be realised except for the fine tuned case of resonant scattering in the early universe with $2 m_\chi\simeq m_V$. We will therefore take $\chi_L$ to be of Majorana type for which the scattering cross section is suppressed by $v_\chi^2 \sim (10^{-3})^2$. Even with such a suppression, the upper bounds from LZ on $g_V/m_V$ can be very strong.
Using the relation between $\epsilon$ and cross section in \cite{Bell:2014tta,Blanco:2019hah},
and  the latest
LZ bound \cite{LZ:2024zvo} which is $\sigma < 5 \times 10^{-48} (m_\chi/200~{\rm GeV})$ cm$^2$ for $m_\chi>200$ GeV, we find
\begin{equation}
	\frac{m_V}{g_V}\gsim  1.7~{\rm TeV}(200~{\rm GeV}/m_\chi)^{1/4}\ . \label{DDS}
\end{equation}
We emphasise again that this bound is derived  from the upper limit on $\epsilon$ which can receive contributions both from the tree level and the loop level, assuming that there is no fine tuned cancellation between them but without any assumption on the exact value of the tree level contribution. That is the reason why instead of "$>$", we  have used "$\gsim$" for the bound.

%%%%%%%%%%%%%%%

Since $\chi_L$ is charged under  the new $U(1)$, its mass requires spontaneous breaking of  $U(1)_{(B-L)_3}$.  As discussed above, in order  to satisfy the bounds from direct dark matter searches, $\chi_L$ should be of Majorana type. The Majorana mass for $\chi_L$ can be obtained after spontaneous gauge symmetry breaking from the following interaction term
$$\frac{\lambda_{\chi\varphi}}{2}\varphi \chi_L^T c \chi_L+{\rm H.c.,}$$
in which $\varphi$ is a new scalar, singlet under the standard gauge group but with  a  $U(1)_{(B-L)_3}$ charge equal to $-2$. $c$ is an antisymmetric $2\times 2$ matrix with nonzero components $\pm 1$ acting on the spinorial indices.
Notice that  even after the gauge symmetry breaking, the $Z_2$ symmetry protecting $\chi_L$ from decay is maintained.
In order to reproduce the CKM mixing matrix, we shall introduce other scalars whose vacuum expectation values (VEV) break  $U(1)_{(B-L)_3}$ but their VEVs are much smaller than $\langle \varphi \rangle$. Thus, the main contribution to the $V$ mass comes from $\langle \varphi\rangle$:
\begin{equation}
	m_V\simeq 2g_V \langle \varphi\rangle.
\end{equation}

The bounds from LHC on searches for the gauge boson of $U(1)_{(B-L)_3}$ for $10~{\rm GeV}<m_V<2m_W$ can be stringent \cite{Elahi:2019drj}. We therefore focus on heavier $V$ where the collider bounds are less stringent. To our best knowledge, there is no dedicated analysis of $U(1)_{(B-L)_3}$ with  the available LHC data; however, there are a number of analyses for simplified models with similar production and decay models for the gauge vector boson \cite{CMS:2025ajo,Fox:2018ldq,Barbosa:2022mmw,CMS:2024nmz}. Combining their results, it seems  that for $m_V>600$ GeV,  $g_V$ can be still as large as $O(1)$ without violating the present collider bounds. However, a signal may be found with a dedicated analysis of the future collider data  if   the  sensitivity to the $V$ production  cross section times the relevant decay branching ratio becomes ten times stronger.
Ref. \cite{Allanach:2026yst} performs a dedicated analysis for the $3B_3-L_1-2L_2$ gauge model and finds that the search for dilepton can constrain the coupling to values  around 0.1. However, since the reconstruction of di-tau ({\it i.e.,} the signature of  our model) is far more complicated than those of  dimuon or an electron positron pair, we expect a dedicated search for the signatures of our model with the present data to  still  allow $g_V$ of order of 1. A dedicated study of the LHC bounds and discovery potential for the $U(1)_{(B-L)_3}$ model is beyond the scope of the present paper.

Since $V$ couples both to $\tau$ and $b$, it can give a new contribution to $\Upsilon\to \tau \bar{\tau}$, leading to
\begin{equation}
	R_{\tau \mu}(\Upsilon)\equiv \frac{\Gamma_{\Upsilon\to \tau \bar{\tau}}/\Gamma_{\Upsilon\to \tau \bar{\tau}}|_{SM}}{\Gamma_{\Upsilon\to \mu \bar{\mu}}/\Gamma_{\Upsilon\to \mu \bar{\mu}}|_{SM}}=\left( 1+ \frac{g_V^2}{e^2}\frac{m_\Upsilon^2}{m_V^2-m_\Upsilon^2}\right)
\end{equation}

According  to \cite{BaBar:2010esv},  $R_{\tau \mu}(\Upsilon (1S))=1.005\pm 0.013 \pm 0.022$.
This measurement implies \cite{Kamada:2018kmi},
\begin{equation}
	\label{Upsilon-bound} \frac{m_V}{g_V} \geq 260~{\rm GeV}
\end{equation}
The value of $R_{\tau \mu}(\Upsilon)$ is also measured for $\Upsilon(2S)$ \cite{CLEO:2006uhx} and $\Upsilon(3S)$ \cite{CLEO:2006uhx,BaBar:2020nlq} with high precision but $\Upsilon(1S)$ provides the strongest bound.
Notice that this bound is far less stringent than the bound from direct dark matter search experiments
shown in Eq. (\ref{DDS}).

The field content of our model along with their quantum numbers are listed in table~\ref{field-content}. We should of course add the vector boson, $V_\mu$ to this list. $\Phi'$ in the last column does not appear in the minimal version of the model described in sect.~\ref{SinglePHI}.
\begin{table}[htbp]
\centering
\begin{tabular}{ccccc}
\toprule
Field & Type & $SU(2)_{\rm EW}$ & $U(1)_{(B-L)_3}$ & $Z_2$\\
\midrule
$\chi_L$ & Majorana & singlet & $1$ & $-1$  \\
$\varphi$ & scalar & singlet & $-2$ & 1\\
$\Phi$ & scalar & doublet & $1/3$ & 1\\
$\phi$ & scalar & singlet & $1/3$ & 1\\
\midrule\midrule
$\Phi'$ & scalar & doublet & $-1/3$ & 1 \\
\bottomrule
\end{tabular}
\caption{Particle content of the model and their quantum numbers}
\label{field-content}
\end{table}

\subsection{Reproducing the CKM mixing matrix}\label{sec:CKM}
In general, the mass matrix of the up and down quarks can be parameterised as
\begin{equation}
	V_{uR}^\dagger \cdot {\rm Diag}[m_u,m_c,m_t]V_{uL} \ \ \ {\rm and} \ \ \ \  V_{dR}^\dagger \cdot {\rm Diag}[m_d,m_s,m_b]V_{dL} \ .
\end{equation}
The CKM mixing matrix can then be written as  $V_{CKM}=V_{uL}V_{dL}^\dagger$. As well-known, the mixing angles  in $V_{CKM}$ are small. Dismissing a fine tuned cancellation between the parameters of $V_{uL}$ and $V_{dL}$ to obtain these small mixing angles, we take the mixing parameters of $V_{uL}$ and $V_{dL}$ to be small and parameterise them as
\begin{eqnarray}
	V_{dL}=\left[ \begin{matrix}
			\cos\theta_d   & \sin \theta_d & \beta_d \cos\theta_d+\alpha_d\sin \theta_d \cr
			-\sin \theta_d & \cos \theta_d & -\beta_d \sin\theta_d+\alpha_d\cos \theta_d \cr
			-\beta_d^*     & -\alpha_d^*   & 1
		\end{matrix}\right] \ {\rm and } \   	V_{uL}=\left[ \begin{matrix}
			\cos\theta_u   & \sin \theta_u & \beta_u\cos \theta_u+\alpha_u \sin \theta_u \cr
			-\sin \theta_u & \cos \theta_u & -\beta_u \sin \theta_u+\alpha_u\cos \theta_u \cr
			-\beta_u^*     & -
			\alpha_u^*     & 1
		\end{matrix}\right] \ . \label{mixingLeft}
\end{eqnarray}

The CKM matrix is then given by \begin{eqnarray} V_{CKM}=V_{u L}V_{dL}^\dagger =\left[ \begin{matrix} \cos(\theta_u-\theta_d)  &  \sin(\theta_u-\theta_d) & -\beta \cos \theta_u-\alpha \sin \theta_u\cr  -\sin(\theta_u-\theta_d) &  \cos(\theta_u-\theta_d) & -\alpha \cos \theta_u+\beta \sin \theta_u \cr \beta^* \cos\theta_d +\alpha^* \sin \theta_d &\alpha^* \cos \theta_d-\beta^* \sin \theta_d & 1\end{matrix} \right]\  \label{CKM}
\end{eqnarray}
in which $\beta=\beta_d-\beta_u$ and $\alpha=\alpha_d-\alpha_u$.
Using the Wolfenstein parameterisation,
\begin{eqnarray} V_{CKM}=\left[ \begin{matrix}
			1-\lambda^2/2            & \lambda     & A\lambda^3(\rho-i\eta)\cr-\lambda & 1-\lambda^2/2 & A\lambda^2\cr
			A\lambda^3(1-\rho-i\eta) & -A\lambda^2 & 1
		\end{matrix} \right]+ \mathcal{O}(\lambda^4)\   \label{Wolf}
\end{eqnarray}
in which the Wolfenstein parameters are  $\lambda =0.22$, $A=0.81$, $\rho=0.117$ and $\eta=0.35$ \cite{ParticleDataGroup:2026}.
Dismissing the possibility of a fine tuned cancellation between $\theta_u$ and $\theta_d$, we find $\cos \theta_u \simeq \cos\theta_d \simeq 1$. Thus, $\beta$ cannot be much larger than $A\lambda^3\sim 8.5\times 10^{-3}$, unless there is a fine tuned cancellation between $\beta$ and $\alpha \sin \theta_d$. Thus, $\alpha \simeq -A\lambda^2$.
One possible solution  \footnote{Eq. (\ref{solution}) is not the only solution. One other alternative is $$ \theta_d\simeq 0   , \  \sin\theta_u\simeq \lambda  \ , \
	\beta\simeq A\lambda^3(1-\rho+i\eta)
	\ {\rm and}  \ \alpha \simeq -A \lambda^2\ .$$
For all possible solutions, $\sin(\theta_u-\theta_d)=\lambda$, $\alpha_u-\alpha_d=A\lambda^2$ and $\beta_{u(d)}/\alpha_{u(d)}\stackrel{<}{\sim}\lambda$. } is
\begin{eqnarray}
	\theta_u\simeq 0   , \  \sin\theta_d\simeq -\lambda  \ , \
	\beta\simeq -A\lambda^3(\rho-i\eta)
	\ {\rm and}  \ \alpha \simeq -A \lambda^2 \ , \label{solution}
\end{eqnarray}

Unlike $V_{dL}$ and $V_{uL}$, $V_{dR}$ and $V_{uR}$  are not directly related to the CKM matrix. We take the general parameterisation for these unitary matrices as
\begin{eqnarray} \label{dRuR}
	V_{dR}^\dagger=\left[ \begin{matrix}
			D_{11} & D_{12} & D_{13}\cr
			D_{21} & D_{22} & D_{23}\cr
			D_{31} & D_{32} & D_{33}
		\end{matrix}\right] \ {\rm and } \   	V_{uR}^\dagger=\left[ \begin{matrix}
			U_{11} & U_{12} & U_{13}\cr
			U_{21} & U_{22} & U_{23}\cr
			U_{31} & U_{32} & U_{33}
		\end{matrix}\right] \ .
\end{eqnarray}

In the mass basis $(\hat{u},\hat{c},\hat{t})$ and
$(\hat{d},\hat{s},\hat{b})$, the gauge interaction of the quarks shown in Eq. (\ref{gV/3})  can be written as follows

\begin{eqnarray} 	\label{dROFF}
	\frac{g_V V_\mu}{3}\left(
	[\bar{\hat{d}}_R \  \bar{\hat{s}}_R \ \bar{\hat{b}}_R]\left[ \begin{matrix} |D_{31}|^2 & D_{31}^*D_{32} & D_{31}^*D_{33}\cr D_{31}D_{32}^* & |D_{32}|^2 & D_{32}^*D_{33}\cr D_{33}^* D_{31} & D_{33}^* D_{32} & |D_{33}|^2 \end{matrix} \right]\gamma^\mu\left[ \begin{matrix}\hat{d}_R \cr  \hat{s}_R\cr  \hat{b}_R\end{matrix}\right]+\right.
\end{eqnarray}
\begin{eqnarray}
	\label{dLOFF}
	[\bar{\hat{d}}_L \  \bar{\hat{s}}_L \ \bar{\hat{b}}_L]\left[ \begin{matrix} 0 & 0 & \beta_{d}\cos\theta_d+\alpha_d\sin\theta_d\cr 0 & |\alpha_{d}|^2 \cos^2\theta_d& \alpha_{d}\cos\theta_d-\beta_d\sin\theta_d\cr \beta_{d}^*\cos\theta_d+\alpha_d^*\sin\theta_d & \alpha_{d}^*\cos\theta_d-\beta^*_d\sin\theta_d & 1 \end{matrix} \right]\gamma^\mu\left[ \begin{matrix}\hat{d}_L \cr  \hat{s}_L\cr  \hat{b}_L\end{matrix}\right]
\end{eqnarray}

\begin{eqnarray} 	\label{uROFF}
	+	[\bar{\hat{u}}_R \  \bar{\hat{c}}_R \ \bar{\hat{t}}_R]\left[ \begin{matrix} |U_{31}|^2 & U_{31}^*U_{32} & U_{31}^*U_{33}\cr U_{31}U_{32}^* & |U_{32}|^2 & U_{32}^*U_{33}\cr U_{33}^* U_{31} & U_{33}^* U_{32} & |U_{33}|^2 \end{matrix} \right]\gamma^\mu\left[ \begin{matrix}\hat{u}_R \cr  \hat{c}_R\cr  \hat{t}_R\end{matrix}\right]+
\end{eqnarray}
\begin{eqnarray} \label{uLOFF}\left.
	[\bar{\hat{u}}_L \  \bar{\hat{c}}_L \ \bar{\hat{t}}_L]\left[ \begin{matrix} 0 & 0& \beta_{u}\cr0& |\alpha_{u}|^2 & \alpha_{u}\cr \beta_{u}^* & \alpha_{u}^* & 1 \end{matrix} \right]\gamma^\mu\left[ \begin{matrix}\hat{u}_L \cr  \hat{c}_L\cr  \hat{t}_L\end{matrix}\right]\right)
\end{eqnarray}
where in Eqs.~(\ref{dLOFF},\ref{uLOFF}), we have neglected higher orders in $\alpha_{u,d}$ and $\beta_{u,d}$ such as $\mathcal{O}(\alpha\beta,\alpha^4,\alpha^2\lambda,\beta^2)$.
Thus, after integrating out $V_\mu$, we shall have effective couplings of the following form in the mass basis of quarks which can induce a contribution to $B^+\to K^+\nu_\tau \bar{\nu}_\tau$  and to $B^+\to K^+\tau \bar\tau$:
\begin{equation}
	\label{GL+GR} \frac{G_L^s+G_R^s}{2}(\bar{\hat{b}}\gamma^\mu \hat{s})(\bar{\nu}_{\tau}\bar{\sigma}_\mu \nu_{\tau}+\bar{\tau}\gamma_\mu \tau )
\end{equation}
in which
\begin{equation}
	\label{GLGR}
	G_L^s=\frac{g_V^2}{3m_V^2}\alpha_d^* \ \ {\rm and } \ \ 	G_R^s=\frac{g_V^2}{3m_V^2}D_{33}^* D_{32}
\end{equation}
where we have neglected terms suppressed by $1-\cos\theta_d$ and $\beta_d \sin \theta_d$.

We will study the effect on $B^+\to K^+\nu_\tau\bar{\nu}_\tau$ in sect. \ref{sec:B+}. Let us however first discuss the bounds on the components of $V_{dR}$ and $V_{uR}$  from various observations.
The strongest bound comes from the $B_s^0-\bar{B}^0_s$ and $B_d^0-\bar{B}^0_d$ mixings. Since the parities of ${B^0}_s$ and
${B^0}_d$ are odd, the main contributions to the mixings come from the axial-axial interaction of the following forms
\begin{equation}
	G_{bs}(\bar{\hat{b}}\gamma^\mu\gamma^5\hat{s})(\bar{\hat{b}}\gamma_\mu\gamma^5\hat{s}) \ \ {\rm and } \ \  	G_{bd}(\bar{\hat{b}}\gamma^\mu\gamma^5\hat{d})(\bar{\hat{b
	}}\gamma_\mu\gamma^5\hat{d})\ .
\end{equation}
In our model,
\begin{equation}
	G_{bs}=\frac{g_V^2}{9m_V^2}\left(\frac{D_{33}^*D_{32}-\alpha_d^*}{2} \right)^2 \ \ {\rm and} \ \  G_{bd}=\frac{g_V^2}{9m_V^2}\left(\frac{D_{33}^*D_{31}-\beta_d^*-\alpha^*\sin\theta_d}{2} \right)^2\ .
\end{equation}

Using the latest bounds  from  the $B_s^0-\bar{B}^0_s$ and $B_d^0-\bar{B}^0_d$ mixings compiled in \cite{Bona:2022zhn,Bona:2024bue} (see also \cite{UTfit:2007eik,Lenz:2010gu}), we find
\begin{equation}
	\frac{g_V| D_{32}^*D_{33}-\alpha_d|}{2m_V}<\frac{1}{165~{\rm TeV}} \label{BMB}
\end{equation}
and
\begin{equation}
	\frac{g_V| D_{31}^*D_{33}+\beta_d+\alpha_d\sin\theta_d|}{2m_V}<\frac{1}{1000~{\rm TeV}}\ . \label{BdMB}
\end{equation}
Eqs. (\ref{dROFF},\ref{uROFF}) show that $V$ also has a tree level coupling to the matter fields, $\hat{d}_R$ and $\hat{u}_R$, given by $g_V|U_{31}|^2$  and $g_V|D_{31}|^2$ on which the bound is not stronger than $\sim 0.1$ for $m_V>700$~GeV \cite{vonAhnen:2021rrf}.
The effective coupling of the induced non-standard interaction for neutrinos,
$$ -\frac{g_V^2}{3m_V^2}\left( |D_{31}|^2 \bar{d}_R\sigma^\mu d_R+ |U_{31}|^2 \bar{u}_R\sigma^\mu u_R\right)(\bar{\nu}_{\tau} \bar{\sigma}_\mu \nu_{\tau}), $$
will be too small to have an observable effect \cite{Farzan:2017xzy}
$$\epsilon_{\tau\tau}\sim \frac{g_V^2(|D_{31}|^2+|U_{31}|^2)/3m_V^2}{\sqrt{2}G_F} \ll 10^{-6}\ .
$$
Similarly, the effect on the dark matter scattering off nuclei due to the coupling suppressed by $|U_{31}|^2$ and $|D_{31}|^2$ will be negligible compared to that induced by the kinetic mixing \cite{Bell:2014tta}
that we discussed before.

\subsection{Minimal version of the model  \label{SinglePHI}}
The mass mixing between the $b$ quark and quarks of the first and second generations requires spontaneous breaking of the $U(1)_{(B-L)_3}$ gauge symmetry. For this purpose, we introduce the  Higgs doublet $\Phi=(\Phi^+ \  \Phi^0)^T$  with a charge of $1/3$ under $U(1)_{(B-L)_3}$.
Then, we can write the following Yukawa couplings
$$
	(Y_{\ell})_\alpha \bar{\ell}_\alpha H^\dagger L_\alpha + Y_{d ij}
	\bar{d}_i H^\dagger Q_j+ Y_{u ij}\bar{u}_i H^T c Q_j +$$
\begin{equation} \label{Phi1}
	\lambda_{t1}  \bar{t}_R \Phi^T c Q_1
	+\lambda_{t2} \bar{t}_R \Phi^T cQ_2+
	\lambda_{d3}  \bar{d}_R \Phi^\dagger  Q_3
	+\lambda_{s3} \bar{s}_R \Phi^\dagger Q_3+{\rm H.c.}
\end{equation}
The mass matrices of the quarks will then take the following form
\begin{eqnarray} \label{YYAsymmetric}[\bar{u}_R \ \bar{c}_R \ \bar{t}_R]\left[ \begin{matrix} Y_{uu}\langle H\rangle & Y_{uc}\langle H\rangle & 0\cr Y_{cu}\langle H\rangle & Y_{cc}\langle H\rangle & 0 \cr \lambda_{t1} \langle \Phi \rangle&  \lambda_{t2} \langle \Phi \rangle& Y_{tt}\langle H\rangle\end{matrix}\right] \left[ \begin{matrix} u_L \cr c_L \cr t_L \end{matrix}\right]=[\bar{u}_R \ \bar{c}_R \ \bar{t}_R]\ V_{uR} ^\dagger \cdot M_u^{diag}\cdot V_{uL}   \left[ \begin{matrix} u_L \cr c_L \cr t_L \end{matrix}\right]\ \end{eqnarray}
and
\begin{eqnarray}
	[\bar{d}_R \ \bar{s}_R \ \bar{b}_R]\left[ \begin{matrix} Y_{dd}\langle H\rangle & Y_{ds}\langle H\rangle & \lambda_{d3} \langle \Phi^* \rangle \cr Y_{sd}\langle H\rangle & Y_{ss}\langle H\rangle &  \lambda_{s3} \langle \Phi^* \rangle\cr 0 & 0& Y_{bb}\langle H\rangle\end{matrix}\right] \left[ \begin{matrix} d_L \cr s_L \cr b_L \end{matrix}\right] =[\bar{d}_R \ \bar{s}_R \ \bar{b}_R] V_{dR} ^\dagger \cdot M_d^{diag}\cdot V_{dL}  \left[ \begin{matrix} d_L \cr s_L \cr b_L \end{matrix}\right]\ \label{bbAsymmetric}
\end{eqnarray}
where $M_d^{diag}$ and $M_u^{diag}$ are diagonal matrices.

Then, Eq. (\ref{YYAsymmetric}) implies
\begin{equation}
	\beta_u^*=-\frac{\lambda_{t1}\langle \Phi\rangle}{m_t}\ ,  \ \alpha_u^*=-\frac{\lambda_{t2}\langle \Phi\rangle}{m_t}\  , \ U_{33}\simeq 1 \ {\rm and} \  U_{31},U_{32},U_{13},U_{23}< \frac{m_c}{m_t}\alpha_u \ll 1. \label{uuU} \end{equation}

Moreover, Eq. (\ref{bbAsymmetric}) gives
\begin{equation} \beta_d=-\alpha_d\sin\theta_d, \ \sin \theta_d=-\frac{Y_{sd}}{Y_{ss}}, \ \alpha_d^*=-\frac{\langle H\rangle \langle \Phi \rangle}{m_b^2}(Y_{ss}\lambda_{s3}^*+Y_{ds}\lambda_{d3}^*),
	\label{ddD}	\end{equation}
$$  D_{13}=\frac{\lambda_{d3} \langle \Phi^*\rangle}{m_b}\ ,  D_{23}=\frac{\lambda_{s3} \langle \Phi^*\rangle}{m_b}\  {\rm and } \ D_{33}=\frac{Y_{b b} \langle H\rangle}{m_b}$$ in which
$$m_b=\left( |Y_{bb}\langle H\rangle |^2+|\lambda_{d3}\langle \Phi\rangle |^2+|\lambda_{s3}\langle \Phi\rangle|^2\right)^{1/2}\  $$ and
$$ D_{11}=\frac{Y_{dd}\langle H\rangle +Y_{ds}\langle H \rangle \sin \theta_d}{m_d} ,  D_{12}=\frac{Y_{ds}\langle H\rangle +\alpha^*_d m_b D_{13}}{m_s} \  {\rm and } \  D_{22}=\frac{Y_{ss}\langle H\rangle +\alpha^*_d m_b D_{23}}{m_s}. $$
Finally,
$$D_{32}=D_{33} \frac{m_b\alpha^*_d}{m_s}  \ \ {\rm and}\ \  D_{31}=-\frac{D_{12}^*D_{32}+D_{13}^*D_{33}}{D_{11}^*}\ .$$

In the absence of a fine tuned cancellation between $\alpha_d$ and $D_{32}^*D_{33}$, Eq.~(\ref{BMB})  implies $\alpha_d$ and $D_{32}^*D_{33}$ to be much smaller than 0.05.  Then, $\alpha_u-\alpha_d=0.04$ (see Eq.~(\ref{solution})) leads to $\alpha_u \simeq 0.04$ so \begin{equation}\langle \Phi\rangle=-\frac{m_t \alpha_u}{\lambda_{t2}}=-7~{\rm GeV}/\lambda_{t2}\ . \end{equation}
Consequently, $|D_{23}|<0.01$ means  $\lambda_{s3}<0.006\lambda_{t2}$.

Since $\Phi$ is charged both under hypercharge and the new gauge symmetry, its VEV can induce a mass mixing between $V$ and the Z boson \cite{Ellis:2018xal}:
$$\delta m^2=\frac{e g_V}{6 \sin \theta_W \cos\theta_W}\langle \Phi\rangle^2\sim (6 ~{\rm GeV})^2g_V\left(\frac{\langle \Phi\rangle}{7 ~{\rm GeV}}\right)^2 \ll  m_Z^2\sin \epsilon $$ so the effect of $\delta m^2$ on the oblique parameters will be much smaller than that of $\epsilon$ and can be neglected. For this,  $\langle \Phi\rangle$ should be much smaller than its mass. Such a small VEV can be obtained by introducing another electroweak singlet scalar charged under $U(1)_{(B-L)_3}$, $\phi$ with the following potential
\be -m_H^2 |H|^2+\frac{\lambda_H}{2} |H|^4-m_\phi^2 |\phi|^2+\frac{\lambda_{\phi}}{2}|\phi|^4+m_\Phi^2 |\Phi|^2+\frac{\lambda_{\Phi}}{2}|\Phi|^4+(A \Phi^\dagger H\phi+{\rm H.c.})
\ee where  we have taken $\lambda_H, \lambda_\phi,\lambda_\Phi>0$ to guarantee being bounded from below. Moreover, we take $A
	\ll m_\Phi$.
This leads to $\langle \phi\rangle\ne 0$, $\langle H\rangle \simeq (m_H^2/\lambda_H)^{1/2}+\delta H$  where $\delta H/ \langle H\rangle =(A^2/8m_\Phi^2 m_H^2)(m_\phi^2/\lambda_\phi)\ll 1$ and
$$ \langle \Phi \rangle= -\frac{A}{2 m_{\Phi}^2} \langle\phi\rangle \langle H\rangle.$$ Taking $A$  small, $\langle \Phi\rangle$ and therefore the contributions from the new physics to the oblique parameters can be made arbitrarily small. $\phi$ will obtain a mixing of $\langle H\rangle A/m_\Phi^2$ with the neutral component of $\Phi$.  Indeed, from the LHC bounds we already know that the components of the electroweak doublet $\Phi$ should be heavier than a few 100 GeV.  If $\Phi$ is lighter than a few TeV, it can be pair produced via electroweak interactions at the LHC and decay to a pair of third generation quark plus a light (first or second generation) quark. A dedicated search may discover such $\Phi$. With the mass of $\Phi$ above $\sim 500$~GeV and $\langle \Phi \rangle \sim 7$~GeV, the typical (physical) mass of $\phi$ will be around 100~GeV. As a result, it will be too heavy to allow $B^+ \to K^+\phi \bar{\phi}$ but $\phi$ can be lighter than $\chi_L$, allowing $\chi_L \bar{\chi}_L \to \phi \bar{\phi}$ for even non-relativistic $\chi_L$ pair. Both $\Phi$ and $\phi$ can reach thermal equilibrium in the early universe.
Then, both $\Phi^0$  and $\phi$  can decay fast in the early universe to $b\bar{s}$ or to  $b\bar{d}$ long before the QCD phase transition era. Notice that the decay of $\phi$ will be possible through the mixing with $\Phi^0$.

\subsection{Symmetric version of the model \label{SymPHI}}
In this section, we shall introduce a model in which $V_{dL} \simeq V_{dR}$ and $V_{uL} \simeq V_{uR}$. As a result, the bounds from $B_s^0-\bar{B}_s^0$ on $\alpha_d$ and $|D_{32}D_{33}|$ can be relaxed, opening the room for explaining the $B^+\to K^+ \nu\bar{\nu}$ excess. For this purpose, we add a new scalar doublet, $\Phi'$ with a $U(1)_{(B-L)_3}$ charge opposite to that of $\Phi$.

$$
	\lambda_{t1}  \bar{t}_R \Phi^T c Q_1
	+\lambda_{t2} \bar{t}_R \Phi^T cQ_2+
	\lambda_{d3}  \bar{d}_R \Phi^\dagger  Q_3
	+\lambda_{s3} \bar{s}_R \Phi^\dagger Q_3+
$$
\begin{equation}
	\lambda_{b1}  \bar{b}_R (\Phi')^\dagger   Q_1
	+\lambda_{b2} \bar{b}_R (\Phi')^\dagger Q_2+
	\lambda_{u3}  \bar{u}_R (\Phi')^T c Q_3
	+\lambda_{c3} \bar{c}_R (\Phi')^T c Q_3+{\rm H.c.}
\end{equation}
%$$ \lambda_{\nu 3} \bar{\nu}_R \Phi_2^T c L_3 +H.c$$
% along with terms such as  $\bar{\tau}_R \Phi^\dagger L_{e,\mu}$ which we drop assuming that the new Higgs, just like the SM Higgs, has a much larger coupling to the top and bottom quarks.
%That is the new interactions maintain the $U_R(2)\times U_L(2)$ quark  flavour symmetry.
%The gauge $U(1)$ symmetry implies that
%$ Y_{u i3}=Y_{u3i}=Y_{d i3}=Y_{d3i}=0$  for $i=1,2$,.

Let us impose the following approximate symmetry to the Lagrangian involving $\Phi$ and $\Phi'$:
\begin{equation}\label{2-3}
	\Phi \leftrightarrow \Phi' \ , Q_3 \leftrightarrow Q_1 \ , s_R \leftrightarrow b_R \ {\rm and} \  c_R \leftrightarrow t_R
\end{equation}
This implies $\lambda_{s3}\simeq \lambda_{b2}$,  $\lambda_{c3}\simeq \lambda_{t2}$ and $\lambda_{t1}\ll \lambda_{t2}$ and    $\lambda_{b1}\ll \lambda_{b2}$. The symmetry is of course broken by the Yukawa couplings of the SM Higgs but to  relax the severe  bound from $B_s^0-\bar{B}_s^0$  on $\alpha_d$ and $D_{32}^* D_{33}$, an approximate symmetry (broken at percent level) will be enough.  

The scalar part of the Lagrangian symmetric under the transformation in Eq. (\ref{2-3})
can be written as
\be -m_H^2 |H|^2+\frac{\lambda_H}{2} |H|^4-m_\phi^2 |\phi|^2+\frac{\lambda_{\phi}}{2}|\varphi|^4+m_\Phi^2 (|\Phi|^2+|\Phi'|^2)+\frac{\lambda_\Phi}{2}(|\Phi|^4+|\Phi'|^4)\ee $$+[A \Phi^\dagger H\phi +(\Phi')^\dagger H\phi^*+H.c.]
$$
Even though we take $m_\Phi^2>0$, $\Phi$ and $\Phi '$ can obtain a VEV suppressed by $\langle H\rangle A/m_\Phi^2$ as
$$ \langle \Phi \rangle=  \langle \Phi' \rangle=-\frac{A}{2 m_{\Phi}^2} \langle\varphi\rangle \langle H\rangle.$$
Then, the quark masses will be approximately equal to
\begin{eqnarray} \label{YY00}[\bar{u}_R \ \bar{c}_R \ \bar{t}_R]\left[ \begin{matrix} Y_{uu}\langle H\rangle & Y_{uc}\langle H\rangle &  \lambda_{t1} \langle \Phi \rangle\cr Y_{cu}\langle H\rangle & Y_{cc}\langle H\rangle &  \lambda_{t2} \langle \Phi \rangle \cr \lambda_{t1} \langle \Phi \rangle&  \lambda_{t2} \langle \Phi \rangle& Y_{tt}\langle H\rangle\end{matrix}\right] \left[ \begin{matrix} u_L \cr c_L \cr t_L \end{matrix}\right]=[\bar{u}_R \ \bar{c}_R \ \bar{t}_R]\ V_{uL} ^T\cdot M_u^{diag}\cdot V_{uL}   \left[ \begin{matrix} u_L \cr c_L \cr t_L \end{matrix}\right]\ \end{eqnarray}
and
\begin{eqnarray}
	[\bar{d}_R \ \bar{s}_R \ \bar{b}_R]\left[ \begin{matrix} Y_{dd}\langle H\rangle & Y_{ds}\langle H\rangle & \lambda_{d 3} \langle \Phi^* \rangle \cr Y_{sd}\langle H\rangle & Y_{ss}\langle H\rangle &  \lambda_{s3} \langle \Phi^* \rangle\cr \lambda_{d 3} \langle \Phi^* \rangle &  \lambda_{s3} \langle \Phi^* \rangle & Y_{bb}\langle H\rangle\end{matrix}\right] \left[ \begin{matrix} d_L \cr s_L \cr b_L \end{matrix}\right] =[\bar{d}_R \ \bar{s}_R \ \bar{b}_R] V_{dL} ^T \cdot M_d^{diag}\cdot V_{dL}  \left[ \begin{matrix} d_L \cr s_L \cr b_L \end{matrix}\right].\label{VdRMVdL}
\end{eqnarray}
The parameterisation of $V_{uL}$ and $V_{dL}$ are given in Eq. (\ref{mixingLeft}). Eqs. (\ref{CKM},\ref{solution}) also apply. Notice that in this symmetric version:
\begin{equation} \label{symm} D_{31}=\beta_d\cos \theta_d+\alpha_d \sin \theta_d,  \ \ D_{32}=\alpha_d\cos\theta_d-\beta_d \sin \theta_d, \ \ {\rm and} \ \  D_{33}\simeq 1
\end{equation}
and
$$ U_{31}=\beta_u\cos \theta_u+\alpha_u \sin \theta_u,  \ \ U_{32}=\alpha_u\cos\theta_u-\beta_u \sin \theta_u, \ \ {\rm and} \ \  U_{33}\simeq 1 \ . $$
Eq. (\ref{symm}) implies that $G_{bs} \ll (G^s_L+G^s_R)^2 m_V^2/g_V^2$ so the contribution of this model to the $B_s^0 -\bar{B}_s^0$ mixing will be negligible but $G^s_L+G^s_R$ can be large enough for explaining the $B^+\to K^+ \nu_\tau \bar{\nu}_\tau$ excess as we shall see in sect. \ref{sec:B+}.

In terms of  the couplings of $\Phi$ and $\Phi'$, the mixing parameters can be written as
\begin{eqnarray}
	\lambda_{t2}\langle \Phi\rangle &=&-m_t \alpha_u^*+m_c(\alpha_u\cos \theta_u-\beta_u \sin \theta_u) \simeq -m_t \alpha_u^* \cr
	\lambda_{t1}\langle \Phi\rangle  &=&-m_t \beta_u^*+m_c\sin \theta_u(\alpha_u\cos \theta_u-\beta_u \sin \theta_u) \simeq  -m_t \beta_u^*
\end{eqnarray}
and
\begin{eqnarray}
	\lambda_{s3}\langle \Phi\rangle &=&-m_b \alpha_d+m_s
	\cos \theta_d(\alpha_d\cos \theta_d-\beta_d \sin \theta_d) \simeq -m_b \alpha_d\cr
	\lambda_{d3}\langle \Phi\rangle  &=&-m_b \beta_d^*+m_s\sin \theta_d(\alpha_d\cos \theta_d-\beta_d \sin \theta_d)\simeq -m_b \beta_d^*
\end{eqnarray}

We can obtain  $\alpha_u \sim \alpha_d\sim O(0.04)$ and $\beta_u\sim \beta_d \sim O(0.001+0.003i)$,  taking $\langle \Phi \rangle\sim 5$~GeV, $\lambda_{t2}\sim 1 $,
$\lambda_{t1}\sim 0.1 $, $\lambda_{s3} \sim m_b/m_t=0.02$ and $\lambda_{d3}\sim 0.1 \lambda_{s3}$.

\section{Explaining the $B^+\to K^++invisibles$ excess\label{sec:B+}}
The Belle~II collaboration \cite{Belle-II:2023esi} has recently reported an evidence for the rare $B^+\to K^+\nu\bar{\nu}$ decay with a branching ratio  $2.7\sigma$ above the SM prediction \cite{Parrott:2022zte,Parrott:2022rgu,Buras:2014fpa}:
\begin{equation}\mathcal{B}(B^+\rightarrow K^+\nu\bar{\nu})
=\left[\,2.3\pm0.5~\text{(stat)}^{+0.5}_{-0.4}~\text{(syst)}\,\right]\times10^{-5}. \label{measurement}\end{equation}
This deviation from the  standard model prediction has motivated a series of papers that try to explain the excess by invoking new physics; see 
\cite{Bause:2023mfe,Bause:2021cna,Ovchynnikov:2023von,Felkl:2023ayn,He:2024iju,Athron:2023hmz,Wang:2023trd,Berezhnoy:2023rxx,He:2023bnk,Datta:2023iln,Altmannshofer:2023hkn,McKeen:2023uzo,Ho:2024cwk,Chen:2024jlj,Loparco:2024olo,Gabrielli:2024wys,Li:2024thq,Hou:2024vyw,Marzocca:2024hua,Rosauro-Alcaraz:2024mvx,Sumensari:2024sji,Hati:2024ppg,Allwicher:2024ncl,Altmannshofer:2024kxb,Abada:2026dlb,Ho:2026kqp,Abumusabh:2026ykp,Prisha:2026nzr,Bhattacharya:2026qzt,Gartner:2026clx,Hong:2026qoj,Yan:2026rws,Bolton:2025lnb,Valencia:2025dte,Kim:2025zaf,Abumusabh:2025zsr,Crivellin:2025qsq,Ding:2025eqq,Aliev:2025hyp,Bolton:2025fsq,Lee:2025jky,Calibbi:2025rpx,Hu:2024mgf,NovoaBrunet:2024axr,Buras:2024mnq}.

%Recent determinations give
%$Br(B^+\to K^+\nu\bar\nu)_{\rm SM}=5.67(38)\times 10^{-6}$ from fully
%relativistic lattice QCD~\cite{Parrott:2022zte,Parrott:2022rgu} where the
%long-distance contribution $Br(B^+\to K^+\nu_{\tau}\bar\nu_{\tau})_{\rm LD}$ was calculated to be  $6.26(55)\times10^{-7}$, and the short-distance part $(4.0\pm 0.5)\times 10^{-6}$ from a combined lattice plus light-cone sum rule
%fit~\cite{Buras:2014fpa}.

Following the standard effective Hamiltonian approach, the SM contribution to the quark-level transition $b\rightarrow s \nu \bar{\nu}$ is given by \cite{Buchalla:1995vs,Altmannshofer:2009ma,%
	Buras:2014fpa}:
\begin{equation} \label{uct}
\frac{G_F}{\sqrt{2}}\,\frac{\alpha}{2\pi\sin^2\theta_W}
	\sum_{l=e,\mu,\tau}
	\Big[
		V_{ub}V_{us}^{*}\,X^l(x_u)
		+V_{cb}V_{cs}^{*}\,X^l(x_c)
		+V_{tb}V_{ts}^{*}\,X^l(x_t)
		\Big]
\end{equation}
where $x_q=m_q^2/m_W^2$ and $X^l(x_q)$ are the Inami--Lim loop functions for the $Z$-penguin and box diagram contributions  (see Ref. \cite{Buchalla:1995vs,Altmannshofer:2009ma,%
	Buras:2014fpa}).
Remembering that $x_u,x_c \to 0$, the GIM mechanism can be used to simplify Eq.~(\ref{uct}) as follows \cite{Buras:1998raa,Misiak:1999yg,Buras:2014fpa}
\begin{equation}
	\begin{aligned}
		\label{eq:SM_Heff_final}
	\frac{G_F}{\sqrt{2}}\frac{\alpha}{2\pi\sin^2\theta_W}
		\sum_{l=e,\mu,\tau}
		V_{ts}^{*}V_{tb}\,X_0(x_t)\,\, {\rm where} \  X_0(x_t)=X^l(x_t)-X^l(0) \ .
	\end{aligned}
\end{equation}

\begin{table}[t]
	\centering
	\begin{tabular}{@{}llll@{}}
		\toprule
		Quantity                         & Value                       &            & Ref.                                 \\
		\midrule
		$m_{B^+}$                        & $5.27941(7)$                & GeV        & \cite{ParticleDataGroup:2026}        \\
		$m_{K^+}$                        & $0.493677(15)$              & GeV        & \cite{ParticleDataGroup:2026}        \\
		$\tau_{B^+}$                     & $1.638(4)\times10^{-12}$    & s          & \cite{ParticleDataGroup:2026}        \\
		$\hbar$                          & $6.582119569\times10^{-25}$ & GeV\,s     & exact                                \\
		\addlinespace[2pt]
		$m_t$                            & $172.60(27)$                & GeV        & \cite{ParticleDataGroup:2026}        \\
		$M_W$                            & $80.3625(77)$               & GeV        & \cite{ParticleDataGroup:2026}        \\
		$\overline{m}_b(\overline{m}_b)$ & $4.186(6)$                  & GeV        & \cite{ParticleDataGroup:2026}        \\
		$\overline{m}_c(\overline{m}_c)$ & $1.2729(45)$                & GeV        & \cite{ParticleDataGroup:2026}        \\
		\midrule
		$|V_{ub}|$                       & $3.89(16)\times10^{-3}$     &            & \cite{ParticleDataGroup:2026}        \\
		$|V_{us}|$                       & $0.22431(85)$               &            & \cite{ParticleDataGroup:2026}        \\
		$|V_{cb}|$                       & $40.7(13)\times10^{-3}$     &            & \cite{ParticleDataGroup:2026}        \\
		$|V_{cs}|$                       & $0.969(5)$                  &            & \cite{ParticleDataGroup:2026}        \\
		$|V_{tb}|$                       & $1.002(24)$                 &            & \cite{ParticleDataGroup:2026}        \\
		$|V_{ts}|$                       & $41.5(9)\times10^{-3}$      &            & \cite{ParticleDataGroup:2026}        \\
		\midrule
		$\eta_{\rm EW}\,G_F$             & $1.1745(23)\times10^{-5}$   & GeV$^{-2}$ & \cite{Parrott:2022zte}               \\
		$\alpha_{\rm EW}^{-1}$           & $127.952(9)$                &            & \cite{Parrott:2022zte}               \\
		$\sin^2\theta_W$                 & $0.23124(4)$                &            & \cite{Parrott:2022zte}               \\
		$X_t$                            & $1.468(17)$                 &            & \cite{Parrott:2022zte,Buras:2014fpa} \\
		\bottomrule
	\end{tabular}
	\caption{Numerical inputs for the $b\to s\nu\bar\nu$ transition. Uncertainties
		are given in parentheses on the last digits.}
	\label{tab:Data-table}
\end{table}
\begin{figure}[htbp]
	\centering
	\includegraphics[width=0.5\linewidth]{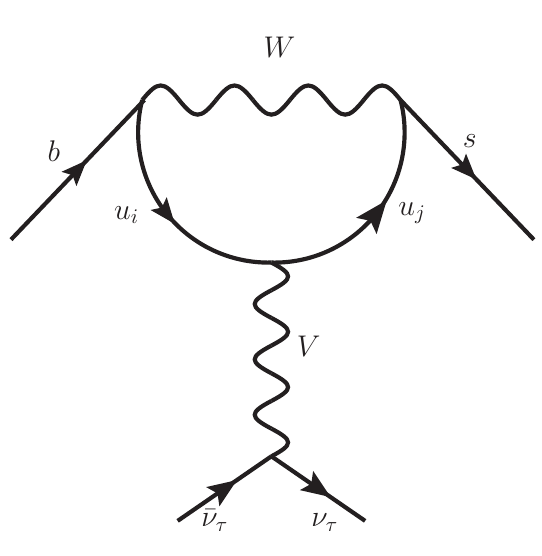}
	\caption{The $V$-penguin contribution to $b\to s\,\nu_\tau\bar\nu_\tau$.}
	\label{fig:new-peng-diag}
\end{figure}
\begin{figure}[htbp]
	\centering
	\includegraphics[width=0.85\linewidth]{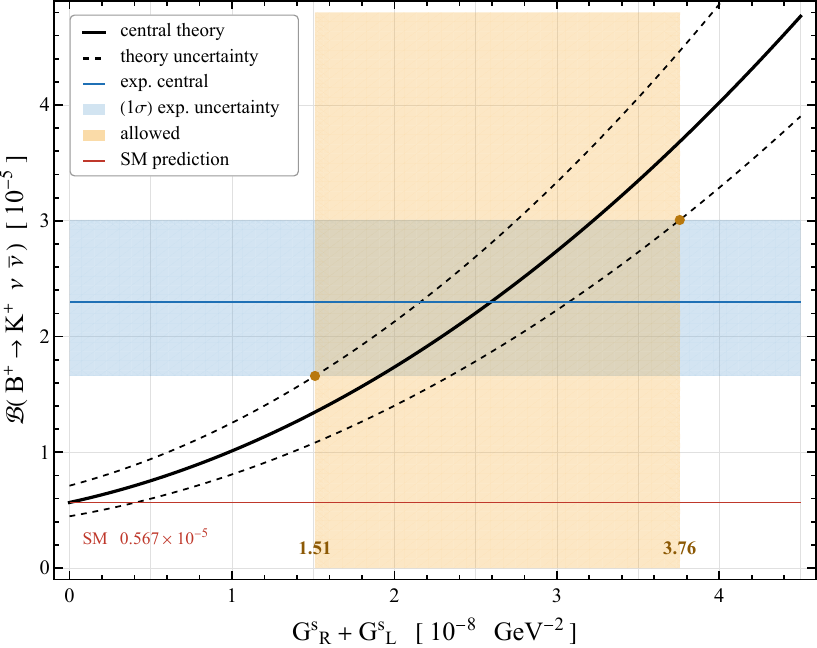}
	\caption{Branching fraction of $B^+\to K^+\nu\bar{\nu}$ as a function of the
		flavour-changing coupling $G_R^s+G_L^s$. The orange highlighted region is the allowed region, obtained from the overlap of the theoretical prediction with the experimental $1 \sigma$ interval.}
	\label{fig:Br_B+_constrains}
\end{figure}
Our model induces  both tree level  and loop level contributions to the amplitude of $B^+\to K^+ \nu_\tau\bar{\nu}_\tau$. The tree level contribution is given by $G_L^s+G_R^s$ in Eqs. (\ref{GL+GR},\ref{GLGR}). The loop diagram is shown in Fig.~\ref{fig:new-peng-diag}.
Both contributions are proportional to $g_V^2/m_V^2$. 
The ratio of the loop level contribution to the tree level is given by
\begin{equation}
	\frac{{\cal M}_{\rm penguin}}{{\cal M}_{\rm tree}}
	=
	\frac{\alpha_{\rm EW}}
	{\sqrt{2}\pi\sin^2\theta_W}
	\frac{|V_{tb}V_{ts}^{*}||X_0^V(x_t)|}{|\alpha_d+D_{33}D_{32}^*|}
	\simeq
	\frac{1.8\times10^{-4}}{|\alpha_d+D_{33}D_{32}^*|}.
\end{equation}
As expected, the loop contribution will be subdominant as long as  $|\alpha_d+D_{33}D_{32}^*|>{\rm few}\times 10^{-4}$. We therefore neglect the loop contribution  and  write
\begin{equation}
	\begin{aligned}
		\frac{d\Gamma(B^+\to K^+ \nu_\tau \nu_{\tau})}{ds}
		 & =
		\frac{m_B^5\,f_+^2(q^2)\,
			\lambda^{3/2}(1,r,s)}
		{1536\,\pi^5\sin^4\theta_W}
		\Big[
			\pi^2\sin^4\theta_W\,(G_L^s+G_R^s)^2
		\\
		 & \qquad
			+2\sqrt{2}\,\alpha_{\text{EW}}\, \eta_{\rm EW}\,G_F\pi\sin^2\theta_W\,
			(G_L^s+G_R^s)\,
			V_{tb}V_{ts}\,X_t
		\\
		 & \qquad
			+2\alpha_{\text{EW}}^2\,(\eta_{\rm EW}\,G_F)^2
			(V_{tb}V_{ts})^2X_t^2
			\Big] ,
	\end{aligned}
	\label{eq:dGamma_BSM}
\end{equation}
where the dimensionless variables are defined as
\begin{equation}
	s\equiv\frac{q^2}{m_B^2},
	\qquad
	r\equiv\frac{m_K^2}{m_B^2},
	\qquad
	\lambda(1,r,s)=1+r^2+s^2-2r-2s-2rs .
\end{equation}
The second term in Eq. (\ref{eq:dGamma_BSM}) is the interference of the SM contribution and that from our model in the tree level. The last term comes from just the SM weak interaction proportional to the square of the Fermi constant, $G_F^2$.
The $\eta_{\rm EW}$ factor   accounts for the universal short distance electroweak corrections to $G_F$ from box diagrams~\cite{Parrott:2022zte}:
\[\eta_{\text{EW}} = 1 + \frac{\alpha_{\text{EW}}}{\pi} \log\left(\frac{M_Z}{M_B}\right) = 1.007(2).
\]
The input values are shown in the Table~\ref{tab:Data-table}.  The non-perturbative QCD effects are embedded in the form factor $f_+(q^2)$. In the appendix, we outline the parameterisation of $f_+(q^2)$.
 
 In our model, $\nu_e$ and $\nu_\mu$ do not  have a new coupling so 
$d\Gamma(B^+\to K^+ \nu_l\bar{\nu}_l)/ds$ for $l=e,\mu$ are given by the SM formula ({\it i.e.,} by Eq.~(\ref{eq:dGamma_BSM}) setting $G_L^s+G_R^s=0$.)
The  rate  of $B^+\to K^++$invisible can be written as 
\begin{equation}
\frac{d\Gamma (B^+\to K^++{\rm invisible})}{ds}=\sum_{l=e,\mu,\tau}\frac{d\Gamma(B^+\to K^+\nu_l\bar{\nu}_l)}{ds}+\frac{d\Gamma (B^+\to {\nu}_\tau \tau^+(\to K^+\bar{\nu}_\tau))}{ds}
	\label{eq:dGamma-tot}
\end{equation}
where the second term on the right-hand side is the famous long distance contribution which is not affected by new physics in our model.

The branching fraction can be obtained after integrating the differential decay rate over the kinematically allowed range of $s$:
\begin{equation}
	\mathcal{B}(B^+\to K^+\nu\bar{\nu})
	=
	\tau_{B^+}
	\int_0^{(1-\sqrt r)^2}
\frac{d\Gamma (B^+\to K^++{\rm invisible})}{ds}  ds.
\end{equation}
The long distance contribution to the branching ratio is given by ~\cite{Parrott:2022zte}:
\[
	\mathcal{B}(B^+\to K^+\nu_\tau\bar{\nu}_\tau)_{\rm LD}
	=
	6.26(55)\times10^{-7}.
\]
The solid black line  in Fig.~\ref{fig:Br_B+_constrains} shows our prediction for $Br(B^+\to K^+ +{\rm invisibles})$ versus $G^s_R+G^s_L$, taking the central values for the input parameters listed in table~\ref{tab:Data-table}.
The main source of uncertainty in the theoretical prediction is due to the form factor $f_+(q^2)$, which is obtained from lattice QCD calculations. This uncertainty is propagated to the branching fraction prediction and is represented by the dashed lines in Fig.~\ref{fig:Br_B+_constrains}.
As shown in the figure  for values of $G_R^s+G_L^s$ in the following range, the observed excess can be explained:
\begin{equation}
	(G^s_R+G^s_L)\in[\,1.5,\,3.8\,]\times10^{-8}~\text{GeV}^{-2}.
\end{equation}

Similarly to $B^+\to K^+\nu_\tau \bar{\nu}_\tau$, there will be a contribution to $B^+\to K^+\tau\bar{\tau}$ at the tree level given by $|G_L^s+G_R^s|^2$.
There are already bounds on this decay mode \cite{Belle-II:2026ism}
\begin{equation}
{\rm Br}(B^+\to K^+\tau^+\tau^-)<0.56\times 10^{-3}
\end{equation}
This bound is too weak to probe the prediction of our model which is $O(10^{-5})$.

Notice that in addition to the off-diagonal coupling of $V$ to $b$ and $s$, the $\lambda_{s3}$ coupling can also convert $b$ to $s$. Let us discuss the possibility of the coupling of $\Phi$ to $\bar{\nu}_R\nu$ to open up a new decay channel $b\to s \bar{\nu}_R\nu$:
$$
 \lambda_\nu \bar{\nu}_R\Phi^Tc L_\tau+m_D\bar{\nu}_L\nu_R$$
 in which the $U(1)_{(B-L)_3}$ charge of $\nu_L$ and $\nu_R$ are both $-2/3$. Since $(\nu_R \ \nu_L)^T$ is taken a Dirac spinor non-chiral under  $U(1)_{(B-L)_3}$, the introduction of the sterile neutrino does not induce gauge anomaly. Moreover, the $m_D$ term will prevent a large mass for $\nu_\tau$ induced by $\lambda_\nu \langle \Phi^0\rangle$ similarly to the  inverse seesaw mechanism. The contribution to ${\rm Br}(B^+ \to K^++{\rm invisibles})$ is expected to be   of order of
 $$ \frac{f_+^2 m_B^5 \lambda_{s3}^2\lambda_\nu^2}{128 \pi^3 m_\Phi^4}\tau_{B^+} \ . $$
As we discuss in sect. \ref{SinglePHI}, within the minimal version of the model with a single $\Phi$, the $B_s^0-\bar{B}_s^0$ mixing implies $\lambda_{s3}<0.006 \lambda_{t2}<0.01$. With this constraint, the contribution from $B^+\to K^+ \bar{\nu}_R\nu_\tau$ can explain the measured excess only with relatively light $\Phi$ of mass around 500~GeV. There are already lower bounds on the $\Phi$ mass from the LHC
\cite{ATLAS:2018gfm}. we will not therefore investigate this line any further and will drop the existence of such sterile neutrinos from our discussion.
\section{Our results \label{sec:DM}}
\begin{figure}[htbp]
	\centering
	\includegraphics[width=0.85\textwidth]{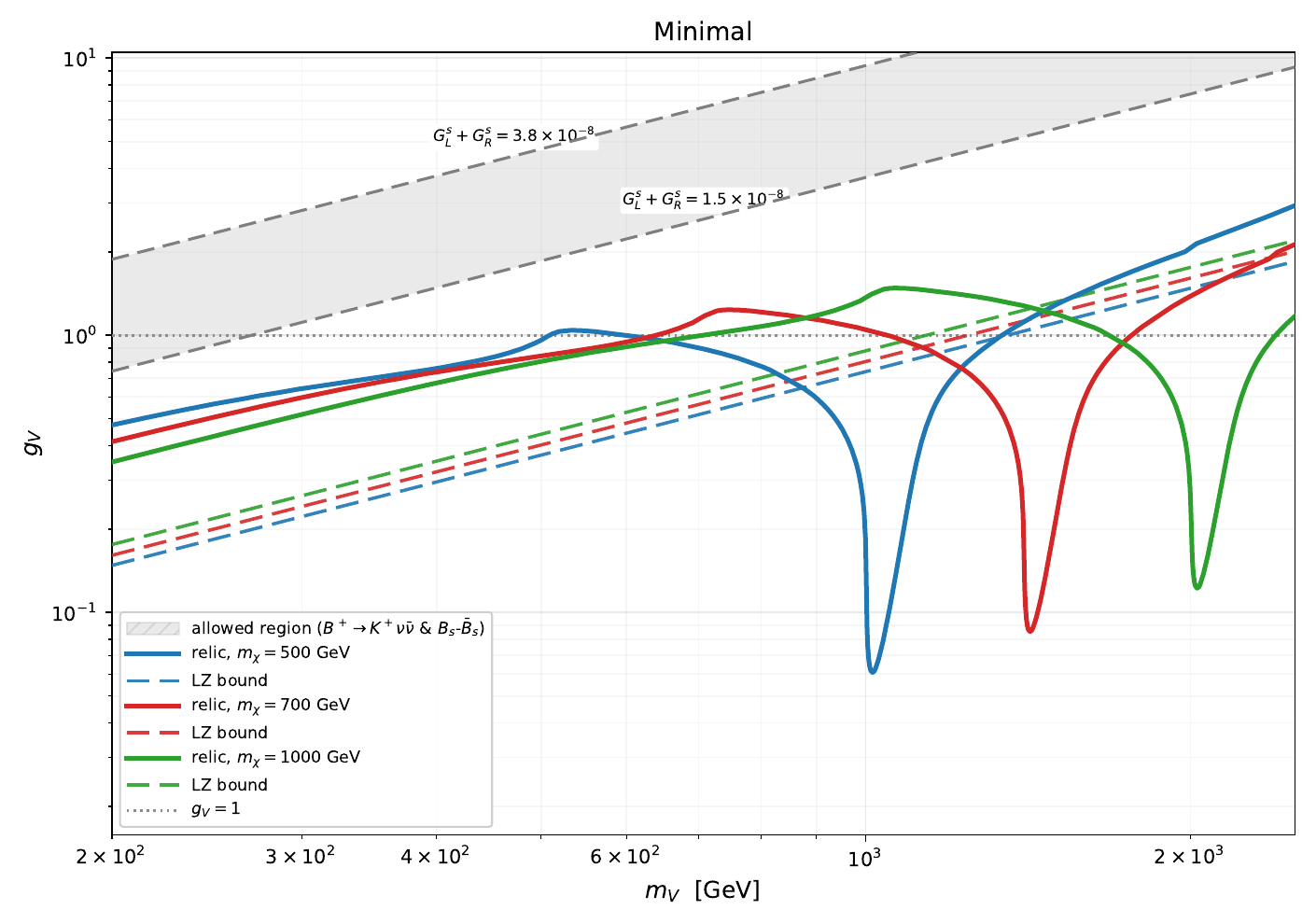}
	\caption{Parameter space of $(m_V,g_V)$ for the minimal scenario. The coloured curves show the values of $g_V$ versus $m_V$ for which the dark matter relic abundance is $\Omega_{DM} h^2=0.12$. The dashed coloured lines show the bound from the LZ direct dark matter detection experiment \cite{LZ:2024zvo}. The blue, red and green lines correspond to $m_\chi=500$~GeV, 700~GeV and 1~TeV, respectively. The grey band shows the range for which the $B^+ \to K^+\nu \bar{\nu}$ excess can be explained, saturating the bound from $B_s^0-\bar{B}_s^0$ in Eq.~(\ref{BMB}) and taking $|\alpha_d|\ll|D_{32} D_{33}|$ or $|\alpha_d|\gg |D_{32} D_{33}|$.}
	\label{fig:minimal}
	
	\vspace{1em}
	
	\includegraphics[width=0.85\textwidth]{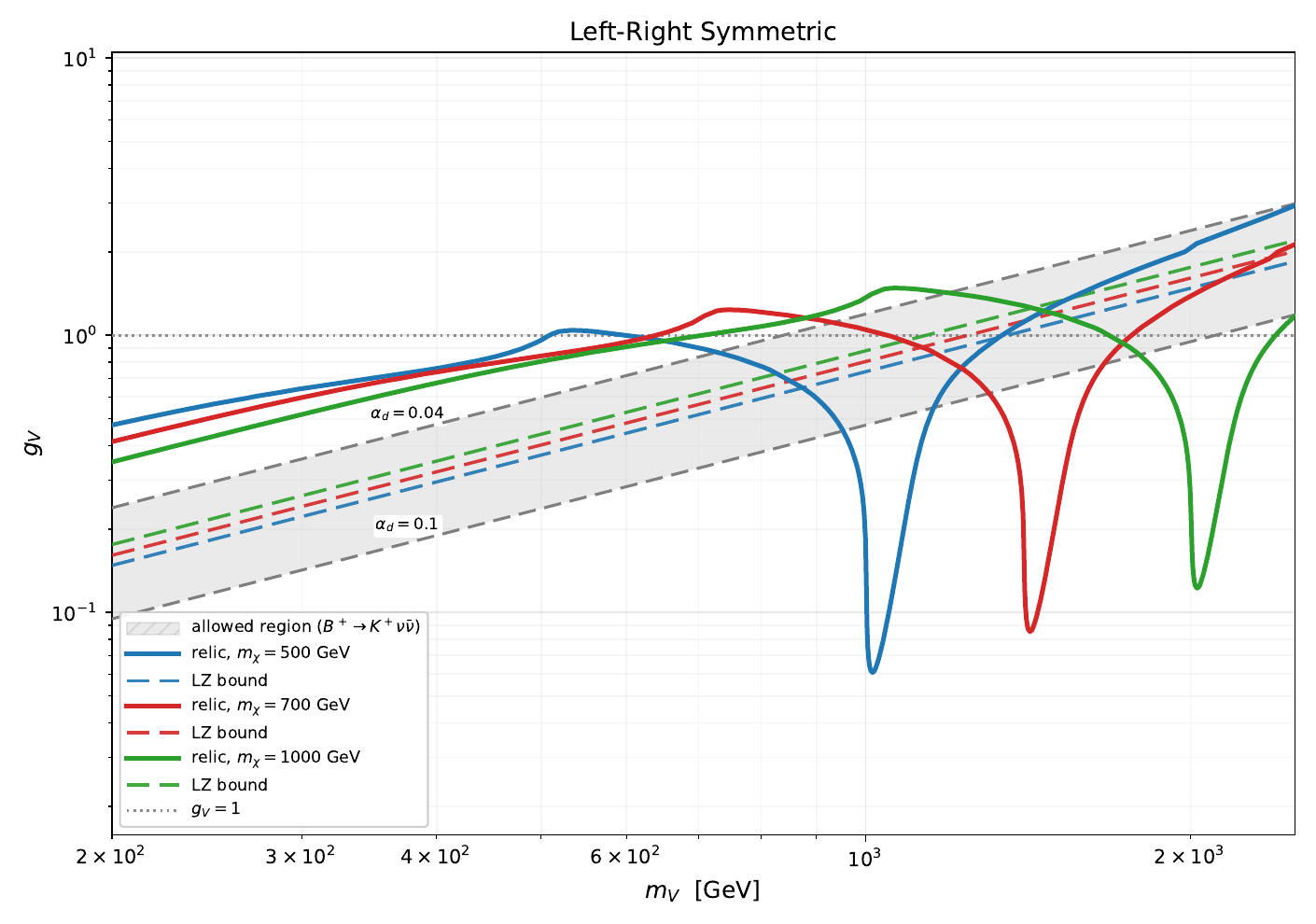}
	\caption{Parameter space of $(m_V,g_V)$ for the left-right symmetric scenario. The coloured solid and dashed lines are the same as those in Fig. \ref{fig:minimal}. The grey band shows the range for which the $B^+ \to K^+\nu \bar{\nu}$ excess can be explained, taking $\alpha_d=D_{32}^* D_{33}$ and varying $\alpha_d$ in $(0.04,0.1)$ (or equivalently in ($-0.1,-0.04)$).}
	\label{fig:left_right_symmetric}
\end{figure}
In our model, the dark matter particles are produced and thermalised  via the new gauge interaction with the third generation fermions of the SM in the early universe. Then, they obtain the observed relic abundance through the freeze-out mechanism via $\bar{\chi}_L\chi_L \to \tau\bar{\tau},\nu_\tau\bar{\nu}_\tau, b \bar{b},t\bar{t}$ and $\phi\bar{\phi}$. We take $\varphi$ and $\Phi$ as well as $\Phi'$ heavier than $\chi_L$ so their presence will not affect
the relic abundance of $\chi_L$. An implication of this assumption is that with  given values of $g_V$ and $m_V$, the relic dark matter abundance for the minimal model with just $\Phi$   will be the same  as that for the left-right symmetric model with both $\Phi$ and $\Phi'$.

Using micrOMEGAs7 \cite{Belanger:2026asz}, we have computed the relic dark matter density. Figs.~\ref{fig:minimal} and~\ref{fig:left_right_symmetric} show $g_V$ versus $m_V$ for $m_\chi=500$~GeV, 700~GeV and 1~TeV with blue, red and green solid curves, respectively. The dashed colour lines show the  LZ upper bound \cite{LZ:2024zvo}
	 corresponding to the mentioned  values of $m_\chi$. At $m_V\simeq 2m_\chi$, the resonant on-shell production of $\chi_L \bar{\chi}_L \to V$ will increase the annihilation cross section for a given $g_V$ which means the measured value of the relic abundance can be obtained with very small $g_V$.  This explains the dips at $m_V\simeq 2 m_\chi$. The direct dark matter bound already rules out $m_\chi\ll m_V/2$ and $m_\chi\gg m_V/2$
but a fine tuned proximity of $m_V$ and $2m_\chi$ is not required for explaining the relic dark matter abundance. 
Remembering that $m_V\simeq 2g_V\langle \varphi\rangle$ and $m_\chi=\lambda_{\chi \varphi} \langle \varphi\rangle$, the proximity of $m_V$ and $2m_\chi$ points towards  $ g_V\sim \lambda_{\chi\varphi}$. It does not  seem unnatural for $g_V$ and $\lambda_{\chi\varphi}$ to be both close to 1. However, the fine tuned relation $m_V\simeq 2m_\chi$ requires some non-trivial explanation for $g_V\simeq \lambda_{\chi \varphi}$.

For $m_V< m_\chi$, $\chi_L \bar{\chi}_L \to VV$ strongly dominates over $\chi_L \bar{\chi}_L \to f \bar{f}$. The output of micrOMEGAs confirms this statement. Notice that unlike $\chi_L \bar{\chi}_L \to f \bar{f}$, the $\chi_L \bar{\chi}_L \to VV$ annihilation can be $s$-wave. However, as seen in Figs.~(\ref{fig:minimal},~\ref{fig:left_right_symmetric}), $m_V<m_\chi$ is already ruled out by the LZ bound.  For $m_V^2 >m_\chi^2 \gg m_t^2$, the dominant annihilation modes are
\begin{equation}{\rm Br}(\chi_L \bar{\chi}_L \to \tau \bar{\tau})\simeq 2{\rm Br}(\chi_L \bar{\chi}_L \to \nu_\tau \bar{\nu}_\tau)\simeq
 3{\rm Br}(\chi_L \bar{\chi}_L\to b \bar{b})\simeq  3{\rm Br}(\chi_L \bar{\chi}_L \to t \bar{t}) \simeq 45 \% \  \label{ann-modes} \end{equation} with  ${\rm Br}(\chi_L \bar{\chi}_L \to \phi \bar{\phi})< 2 \%$.
The relations in Eq.(\ref{ann-modes}) can be easily understood as  follows. Since the $U(1)_{(B-L)_3}$ charges of $\nu_\tau$ and $\tau$ are the same, we expect $
{\rm Br}(\chi_L \bar{\chi}_L \to \tau_L \bar{\tau}_L)\simeq {\rm Br}(\chi_L \bar{\chi}_L \to \tau_R \bar{\tau}_R)\simeq  {\rm Br}(\chi_L \bar{\chi}_L \to \nu_\tau \bar{\nu}_\tau)
$ up to $\mathcal{O}(m_\tau^2/m_\chi^2)$ correction. Summing over the helicities of $\tau$, we arrive at the first relation in Eq.~(\ref{ann-modes}). On the other hand, the 
$U(1)_{(B-L)_3}$ charges of  $t$ and $b$ are $1/3$ of that of $\tau$ so the cross section of annihilation to each colour of $t$ or $b$ will be suppressed by $(1/3)^2$. Counting the final colours, we arrive at the last two equations in Eq.~(\ref{ann-modes}).
We have also examined the case in which $\chi_L \bar{\chi}_L \to \phi \bar{\phi}$ is kinematically forbidden. As expected the shift in the values of $g_V$ that give the observed abundance is small and below 1 \%. That is the inclusion of $\phi$ does not change the overall picture.

The grey bands in Fig.~\ref{fig:minimal} and Fig.~\ref{fig:left_right_symmetric} show the range that can explain the $B^+\to K^++{\rm invisibles}$ excess, respectively within the minimal and left-right symmetric versions of our model.
For the minimal version, we use the bound from the $B_s^0-\bar{B}_s^0$ mixing formulated in Eq.~(\ref{BMB}) and equate $G_L^s+G_R^s$ with $(2/165~{\rm TeV})g_V/m_V$ which is based on two assumptions: (1) The bound in Eq.~(\ref{BMB}) is saturated; (2)  Either $|\alpha_d|\ll |D_{32}D_{33}|$ or $|\alpha_d|\gg |D_{32}D_{33}|$. The lower and upper limits of the grey band in Fig.~\ref{fig:minimal} respectively correspond to $G_L^s+G_L^s=1.5\times 10^{-8}$ GeV$^{-2}$ and  $3.8\times 10^{-8}$ GeV$^{-2}$.
To draw the grey band in Fig.~\ref{fig:left_right_symmetric}, we have taken $\alpha_d=D_{32}^*D_{33}$. Notice that $B^+ \to K^* \nu_\tau\bar{\nu}_\tau$ is not sensitive to the sign of  $G_L^s+G_R^s$ and therefore to that of $\alpha_d$. The lower and upper limits of the grey band
 respectively correspond to $\alpha_d=\pm 0.1$ and $\alpha_d=\pm 0.04$. Remembering that $\alpha=-0.04=\alpha_d-\alpha_u$, these values of $\alpha_d$ do not require a fine tuned cancellation between $\alpha_u$ and $\alpha_d$.

As seen in Fig.~\ref{fig:minimal},  explaining the $B^+\to K^+ \nu \bar{\nu}$  excess within the minimal version of our model requires such large values of $g_V$ which will lead to under-abundance of relic $\chi_L$. Even if we forget the  connection to the dark matter, explaining the excess within the $U(1)_{(B-L)_3}$ model is still challenging because for $g_V<1$, the $V$ mass should be lighter than 300 GeV which may have already been ruled out by the LHC direct search for $V_\mu$ \cite{Elahi:2019drj}. However, in the left-right symmetric version where the bound from  $B_s^0-\bar{B}_s^0$ 
 is avoided, there is ample possibility for explaining the $B^+\to K^+\nu \bar{\nu}$ excess and  still obtaining the observed relic abundance, without violating the direct dark matter detection bounds. This typically requires $800~{\rm GeV}<m_V<2~{\rm TeV}$ with $0.2<g_V<1$ which may be within the discovery reach of the upcoming run of the LHC. A dedicated study, beyond the scope of the present paper, is required for this purpose.
 \section{Conclusions \label{sec:Con}}
We have studied the $U(1)_{(B-L)_3}$ gauge model in which the chiral fermion, $\chi_L$, added to cancel the $[U(1)_{(B-L)_3}]^3$ anomaly, plays the role of the dark matter. Through this gauge interaction, the dark matter can be produced and thermalised in the early universe. Its relic abundance is determined via the $s$-channel annihilation to $\nu_\tau\bar{\nu}_\tau, \tau \bar{\tau},t\bar{t}$ and $b\bar{b}$. The third generation fermions charged both under $U(1)_{(B-L)_3}$ and the  electroweak $SU(2)\times U(1)_Y$  can induce kinetic mixing between the new  and the hypercharge gauge bosons. For the gauge boson mass below $\sim 200$ GeV, the constraint from the electroweak precision data on this mixing  is rather  weak and around 0.1 \cite{Ellis:2018xal}. Such a kinetic mixing would however lead to a signal in the direct dark matter search experiments, unless the dark matter is of Majorana type with a cross section suppressed by the square of its velocity.
Even with this suppression, the LZ bound  \cite{LZ:2024zvo} restricts the parameter space, pointing towards $\chi_L$ and $V$ heavier than 500~GeV. Lighter dark matter is possible only at the resonance for which $m_V \simeq 2 m_\chi$ so that an efficient resonant $\chi_L \bar{\chi}_L$ annihilation can take place with very small gauge coupling. We have focused on $m_V>200$ GeV and $m_\chi>500$~GeV. The LHC searches for $V$ as well as the electroweak precision data  point towards such heavy $V$, too \cite{Ellis:2018xal}.

Since the dark matter in this model is of Majorana type, the annihilation will dominantly be $p$-wave and suppressed with the square of the velocity of the annihilating pair relative to each other. As a result, the annihilation cross section in the freeze-out epoch  with $v_\chi\sim 0.3$ will be far larger than that in the galaxy with $v_\chi \sim 10^{-3}$ so we do not expect a conventional indirect dark matter signal. However, an indirect signal may come from dark matter spike around supermassive black hole in the centre of Milky Way where  the dark matter density and velocity can be very large \cite{Gondolo:1999ef,Balaji:2023hmy}.

Within our model, on one hand, we expect that with a slightly larger exposure of   the direct dark matter search experiments, the signal of a heavy dark matter with mass larger than $\sim 500$ GeV to be discovered but on the other hand, we expect null results from the indirect dark matter searches in dwarf galaxies,  in DM halo and  in galaxy clusters. If the future measurements confirm these predictions and high luminosity LHC discovers a new gauge boson coupled to the third generation fermions, it will still remain unknown whether the discovered dark matter  is also charged under this new gauge symmetry or not, especially for $m_V<2 m_\chi$ for which $V$ does not decay to a $\chi_L$ pair. A positive photon (neutrino) signal from the annihilation of a DM pair into $t\bar{t}$, $\tau \bar{\tau}$ and $b\bar{b}$ (into $\nu_\tau \bar{\nu}_\tau$) in the  dark matter spike around the central supermassive black hole of the Milky Way  will then  be a strong hint in favour of our model for the dark matter.

Since the $U(1)_{(B-L)_3}$ charge of the $b$ quark is different from that of $s$ (and $d$), their mixing requires gauge symmetry breaking. We   have introduced two variants of the model in which such a mixing can be obtained: (1) In the first variant,  the electroweak Higgs doublet with a VEV of $\sim 7$ GeV is responsible for the mixing between the third generation quarks with the first and second generations. We show that the full CKM matrix can be accommodated in this minimal version. The quark mass matrices will not be symmetric so the mixing matrices of right-handed and left-handed quarks will not be the same. Going to the mass basis of the quarks, this means FCNC currents  both of vector form ({\it i.e.,} $\bar{s}\gamma^\mu b V_\mu$ or $\bar{d}\gamma^\mu b V_\mu$) and of axial form ({\it i.e.,} $\bar{s}\gamma^\mu  \gamma^5 b V_\mu$ or $\bar{d}\gamma^\mu \gamma^5 b V_\mu$) can be obtained.  On the axial FCNC couplings ($\bar{s}\gamma^\mu  \gamma^5 b V_\mu$ or $\bar{d}\gamma^\mu \gamma^5 b V_\mu$), there is a strong bound from the $B^0_s-\bar{B}^0_s$ mixing which prevents a contribution to $B^+ \to K^+ \nu_\tau \bar{\nu}_\tau$ large enough to explain the excess reported by Belle~II.
(2) In the second variant, we introduce two doublets, $\Phi$ and $\Phi'$ with a Lagrangian symmetric under $\Phi \leftrightarrow \Phi'$ which predicts a symmetric  mass matrix for the quarks. Then, in the mass basis, while the vector coupling of form  $\bar{b} \gamma^\mu s V_\mu$ is obtained, the axial coupling, $\bar{b} \gamma^\mu \gamma^5 s V_\mu$ disappears. As a result, we can have  a relatively large contribution to $B^+\to K^+\nu_\tau \bar{\nu}_\tau$ at the tree level explaining the reported excess. 
Similarly, a contribution of $10^{-5}$ to ${\rm Br}(B^+ \to K^+\tau^+\tau^-)$ is expected which is beyond the resolution of the present measurements \cite{Belle-II:2026ism}.
$\Phi$ and $\Phi'$ can be both heavier than $\sim 500$ GeV, escaping the present LHC bound but can in principle be pair produced in the colliders via their electroweak interactions and promptly decay to two jets of third and second (or first) generations. We have introduced a mechanism for $\Phi$ and $\Phi'$ to obtain a small and equal VEV of $\sim 5$ GeV, despite their large masses.

%%%%%%%%%%%%%%%%%%%%%%%%
%%%%%%%%%%%%%%%%%%%%%%%%%%\\
%%%%%%%%%%%%%%%%%%%%%%
%%%%%%%%%%%%%%%%%%%%%%%%%%%

\section*{Appendix: parameterisation of form factor, $f_+(q^2)$} \label{app:form}
 We use the results  from the lattice QCD calculation in Ref.~\cite{Parrott:2022rgu} to parametrise the form factor $f_+(q^2)$. The parameterisation is based on the $z$-expansion as:
\begin{equation}
	\begin{aligned}
		f_+(q^2, & M_H)  =
		\frac{\mathcal{L}^{\mathrm{cont}}(M_H)}
		{1-{q^2}/{M_{B_s^*}^2}}
		\sum_{n=0}^{N-1}
		a_n^{+,\mathrm{(cont)}}(M_H)
		\left(
		z(q^2)^n
		-
		\frac{n}{N}(-1)^{\,n-N}z(q^2)^N
		\right)                         \\[6pt]
		& = \frac{\mathcal{L}}
		{1-{q^2}/{M_{B_s^*}^2}}
		\Bigg[
		a_0^{+}
		+ a_1^{+}\left(z(q^2)-\frac{1}{3}z(q^2)^3\right)
		+ a_2^{+}\left(z(q^2)^2+\frac{2}{3}z(q^2)^3\right)
		\Bigg] \ ,
	\end{aligned}
\end{equation}
where 
\begin{equation}
	\begin{aligned}
		z(q^2,t_0) & =
		\frac{
			\sqrt{t_+ - q^2}
			-
			\sqrt{t_+ - t_0}
		}{
			\sqrt{t_+ - q^2}
			+
			\sqrt{t_+ - t_0}
		}
	\end{aligned}
\end{equation}
in which
\begin{equation}
		t_+  = (M_B + M_K)^2
\  , \ 
		t_-  = (M_B - M_K)^2 \ {\rm and} \ 	t_0 = 0 \ .
\end{equation}

The numerical values for the $z$-expansion are presented in Table~\ref{tab:fits}.
\begin{table}[htbp]
	\centering
	\begin{tabular}{lc}
		\toprule
		Parameter      & Value                     \\
		\midrule
		$\mathcal{L}$  & $1.304(10)$               \\
		$a_0^{0}$      & $0.2545(90)$              \\
		$a_1^{0}$      & $0.210(76)$               \\
		$a_2^{0}$      & $0.02(17)$                \\
		$a_0^{+}$      & $0.2545(90)$              \\
		$a_1^{+}$      & $-0.71(14)$               \\
		$a_2^{+}$      & $0.32(59)$                \\
		$M_{B_s^*}$    & $5.4158(15)\ {\rm GeV}$   \\
		$M_{B_{s0}^*}$ & $5.729495(85)\ {\rm GeV}$ \\
		\bottomrule
	\end{tabular}
	\caption{Parameters of the $z$-expansion fit for the $B\to K$ form factors,
		taken from Ref.~\cite{Parrott:2022rgu}.}
	\label{tab:fits}
\end{table}
\acknowledgments
Authors acknowledge K. Azizi for collaboration in the  early stages of the project.
 M. A. thanks Saeed Abbaslu for assistance with micrOMEGAs.
M.A. acknowledges the school of  theoretical physics, Institute for Research in Fundamental Sciences (IPM) for the partial financial support, office facilities and the hospitality of its staff. 
Y. F.  is grateful to  IFIC, University of Valencia where this work was initiated.  
This project has received funding from the European Union’s Horizon Europe research and innovation programme under the Marie Skłodowska-Curie Staff Exchange grant agreement No 101086085 – ASYMMETRY.
\bibliographystyle{unsrtnat}
\bibliography{bibTOP}

\end{document}